# Self-healing in unpassivated and passivated CdTe nanostructures: structural stability and optical properties

Kashinath T Chavan, Sharat Chandra

*Materials Science Group, Indira Gandhi Centre for Atomic Research, Kalpakkam 603102, Tamil Nadu, India*
*Homi Bhabha National Institute, Training School Complex, Anushakti Nagar, Mumbai, 400094, India*
email: ktchavan99@gmail.com, sharat@igcar.gov.in

## Abstract

We report the effects of passivation on the various properties like electronic structure, structural stability and optical properties of *CdTe* in the different nanostructure forms such as ultra-thin slabs, monolayers, nanorods and nanotubes. Further, based on these properties, the self-healing ability of each nanostructure has been predicted. The optical properties suggest that all of the passivated and specific unpassivated nanostructures are suitable for optoelectronic applications. The 2D system in <110> orientation and nanotube derived from the <111> monolayer show significant self-healing in the pristine structures.

## 1. Introduction

The discovery of graphene and its derivatives has shown the vast importance of such single-atom thick monolayers in nanotechnology. Motivating from this, similar nanostructures of different materials have been investigated for technological applications [1-3]. Various low-dimensional nanostructures (slabs, monolayers, quantum dots, atomic clusters, nanorods, nanotubes, self-assembled nanostructures, etc.) have been investigated for their remarkable optical properties.

Since the material properties in low dimensions are dictated by the surface effects and quantum confinement, it does not necessarily match with its bulk properties. The unsatisfied bonds at the surface and the almost infinite potential experienced by surface electrons contribute to the surface effects. However, the surface effects can be partially neutralized by surface passivation using an appropriate ligand such that the dangling bonds are satisfied. The passivation method is chosen depending on the nanostructure's geometry and chemistry. For example, the grain boundaries of 3D grains can be passivated using 2D layers [4].

The II-VI compound semiconductor, cadmium telluride (*CdTe*) has a direct energy gap of ~1.54 eV and is an essential material for optoelectronics, radiation detectors and spintronics. This direct energy gap of 1.54 eV and the high absorption coefficient makes *CdTe* a candidate for the absorber material in the solar cell. The Hg and Zn doping makes it into IR and gamma-ray detector [5], respectively, and upon doping with magnetic impurity atoms, it can acquire a magnetic and half-metallic nature [6]. The monolayer of *CdTe* is also proposed as a photocatalyst in the water splitting applications [7]. *CdTe* can be easily converted from intrinsic to extrinsic semiconductor by controlling its stoichiometry, into p-type (Te rich) or n-type (Cd rich) [8] material.

The effect of quantum confinement is observed when the material's dimensions are of the order of excitonic Bohr radius ($\alpha_B$). The excitonic Bohr radius is intrinsic to the semiconducting material. The degree of confinement is decided by comparing the system size to its excitonic Bohr radius. Therefore, depending on the system size (*a*), the effect of confinement is classified into weak ($a \sim \alpha_B$), moderate ($a < \alpha_B$) and strong ($a \ll \alpha_B$) [9]. The excitonic Bohr radius for the *CdTe* is reported to be in the range of ~65 Å - 75 Å [10-12], and it is the largest among the IIB-VI group but it has smallest exciton binding energy which can lead to unstable excitonic emission [10]. However, its excitonic binding energy increases in nanocrystalline form [9].

For solar cell applications, the ultra-thin *CdTe* films have lower costs and high throughput [13]. Recent studies by Wu *et al*. have reported stretchable photodetectors using the ultrathin *CdTe* films of a few nanometer thickness [14]. The 2D monolayers with tetragonal crystal structures of IIB-VI compounds have been predicted as the photocatalyst for water splitting applications [7]. Unsal *et al*. have studied the monolayer of *CdTe* with an energy gap of 1.95 eV for the optoelectronic device applications, and this energy gap is dependent on the strain in the monolayer [15]. Buckled hexagonal monolayer of *CdTe* has been studied theoretically for electronic and spin transport and predicted to be a potential candidate for flexible piezospintronic devices [16]. Ultra-thin films of *CdTe* with a thickness of 4.6 nm have been synthesized for optoelectronic applications [14,17]. The developments in technology are enabling the synthesis of few-layer thick *CdTe* films, therefore, the synthesis of monolayers could also be possible soon as reports suggest their stability [15, 18-20].

Apart from the 2D *CdTe* nanostructures, the 1D nanostructures like, nanorod, nanotube are also being studied [21-22]. Hollow nanostructures like nanotubes can be synthesized from the core-shell nanowires using the Kirkendall growth mechanism [23]. In nanotubes, the confinement effect becomes more profound as the diameter of tubes approaches the excitonic Bohr radius [24]. The quantum confinement in the *CdTe* nanotubes results in strong luminescence in the visible region, as seen in photoluminescence and Raman spectroscopy [22]. Sarkar *et al*. have studied the double walled *CdTe* nanotubes for their band gap engineering [25]. The *Cd-Te* bond lengths in nanotubes are shorter than that in nanowires due to the dominance of $sp^2$ hybridization in nanotubes. The passivation of surface atoms results in the *Cd-Te* bond lengths close to that of bulk, and thus the passivation lowers the surface strain [25].

The development of self-healing semiconducting materials or nanostructures is essential for device applications. In nature, the injuries of living organisms get repaired on their own, which is called self-healing [26]. Similarly, when nanostructures are cleaved from bulk, bonds of atoms on the new surfaces are broken, leading to high chemical reactivity and thus instability. But in many instances, the surface undergoes reconstruction and bonds of surface atoms re-orient to satisfy their valance, leading to stability and reduced chemical reactivity. Such materials do not need surface passivation and are helpful in device applications. Usually, the intrinsically self-healing materials have low elastic modulus (soft) and are deformable [27]. Orellana *et al*. report enhancement of the self-healing in the *CdTe* structures on chlorine doping [28]. Despite extensive research that has been carried out for *CdTe* nanostructures, to the best of our knowledge, there are no reports of single-walled nanotubes of *CdTe*. In literature, we find few papers discussing the monolayers of *CdTe* and the effect of surface passivation on the properties of 1D and 2D nanostructures of *CdTe* have not yet been reported. The comparison of properties of pristine and passivated nanostructures can help us in understanding the self-healing ability of these nanostructures.

This paper presents the structural and dynamic stability, electronic structure and optical properties of *CdTe* in different nanostructural forms such as the thin slab, monolayer, nanorod and nanotube. We study the effect of passivation on various properties for the analysis of self-healing ability of pristine nanostructures. Further, we discuss the possible applications of *CdTe* ultra-thin 2D layers, 1D nanorod and nanotubes.

## 2. Calculation Details

The geometry relaxation, electronic structure, optical properties, *ab-initio* molecular dynamics (AIMD), phonon dispersions, etc., are calculated using the first-principles density functional theory as implemented in the VASP [29-33]. The exchange-correlation functional is approximated within the generalized gradient approximation and PAW PBE scheme used for the pseudopotentials [34-36]. All the structures are relaxed to the maximum force tolerance of $10^{-3}$ eV/Å, and for self-consistency, $10^{-7}$ eV is used as the total energy convergence criterion. A minimum vacuum of 25 Å has been kept in the respective confinement directions to minimize the system's interaction with its periodic image. For electronic structure calculations, the kinetic energy cutoff used for the plane waves is 700 eV for the pristine and 800 eV for passivated nanostructures of *CdTe*. The K point grid and kinetic energy cutoff for plane waves used in the calculations of the electronic structure of *CdTe* nanostructures are given in table ST1 of supporting information (SI). The cleaved or rolled *CdTe* nanostructures are relaxed to their nearest local minima using the conjugate-gradient method [37]. Since the *CdTe* has a zinc blende structure with tetrahedral bonding, the charge of each atom is shared by the four other atoms. Thus, the *Cd* with a valance 2 ($4d^{10}5s^2$) shares 0.5 electronic charge with each *Te*, whereas the *Te* has valence 6 ($5s^25p^4$), and it shares 1.5 electronic charge with each *Cd* atom. Due to the bond-breaking, the nanostructures have dangling bonds on their surface atoms due to unsatisfied valance. Therefore, to satisfy their valance, the surface atoms are passivated using fictitious hydrogen-like pseudo atoms (FHPA) with a fractional atomic number. For the passivation of *Cd*, FHPA with 1.5 atomic number, whereas for *Te*, FHPA with 0.5 atomic number are used [38] and this quenches the chemical reactivity of surface atoms. The AIMD and phonon dispersion calculations are carried out using the 600 eV energy cutoff for plane waves. The nanostructures were subjected to AIMD to study their structural stability at 300 K. For that the structures are heated slowly from 0 K to 300 K within the micro-canonical ensemble and 1 ps simulation time. Further, a constant temperature MD is performed for equilibration within the canonical ensemble for more than 10 ps.

To derive the tube, the rectangular monolayers are rolled up using the following formulae.

$X = R\cos\theta, Y = R\sin\theta$, $R = L/2\pi$ and $\theta = (\frac{L}{2\pi} + \delta)/(\frac{L}{2\pi})$ in radians,

where, $L$ is the total length of layer along the direction to be rolled, $\theta$ is the angle of rotation and $\delta$ is the linear distance from the origin. The per atom binding energies are calculated using the following the formula

$BE = [\{NE(Cd) + ME(Te)\} - E(Cd_N Te_M)]/(N + M)$,

where the first term in numerator is addition of total energies of isolated Cd and Te atoms and the second term i.e., $E(Cd_N Te_M)$ is that of nanostructure.

## 3. Results and discussion

### 3.1 SLABS

The thin *CdTe* slabs in pristine and surface passivated form are studied and their geometries are shown in figure 1. The <100> and <110> oriented slabs comprise 2 monolayers whereas <111> oriented slab has 3 monolayers. The structural stability of these structures has been studied using the AIMD at 300 K and their energy vs. time profile is shown in figure S1 in SI. When these structures are heated to 300 K and subjected to constant temperature AIMD, the pristine slabs in <100> and <111> undergo surface reconstruction at 300 K and thermalize at a lower average energy. These new structures with a lower energy for <100> and <111> slab are relaxed to the nearest local minima at 0 K and are used for further calculations. The pristine <110> slab and all the passivated forms of slabs, retain their initial shape and are intact at 300 K and thus are structurally stable.

The passivation of slabs using FHPA results in the change in its geometry and changes are measured in the form of slab thickness and *Cd-Te* bond lengths. The slab thickness increases from 5.14 Å to 6.55 Å for <100>, 8.42 Å to 8.51 Å for <111> whereas it decreases from 7.5 Å to 6.95 Å for <110> oriented slabs upon passivation.

The pristine structures have the *Cd-Te* bond lengths over quite wide range which reduces to over the short range on passivation. The range of *Cd-Te* bond lengths in pristine cases are 2.79 Å - 3.15 Å (<100>), 2.81 Å - 2.93 Å (<110>), 2.70 Å - 3.28 Å (<111>) whereas the passivated cases have 2.86 Å - 2.88 Å (<100>), 2.83 Å - 2.87 Å (<110>), 2.85 Å - 2.87 Å (<111>).

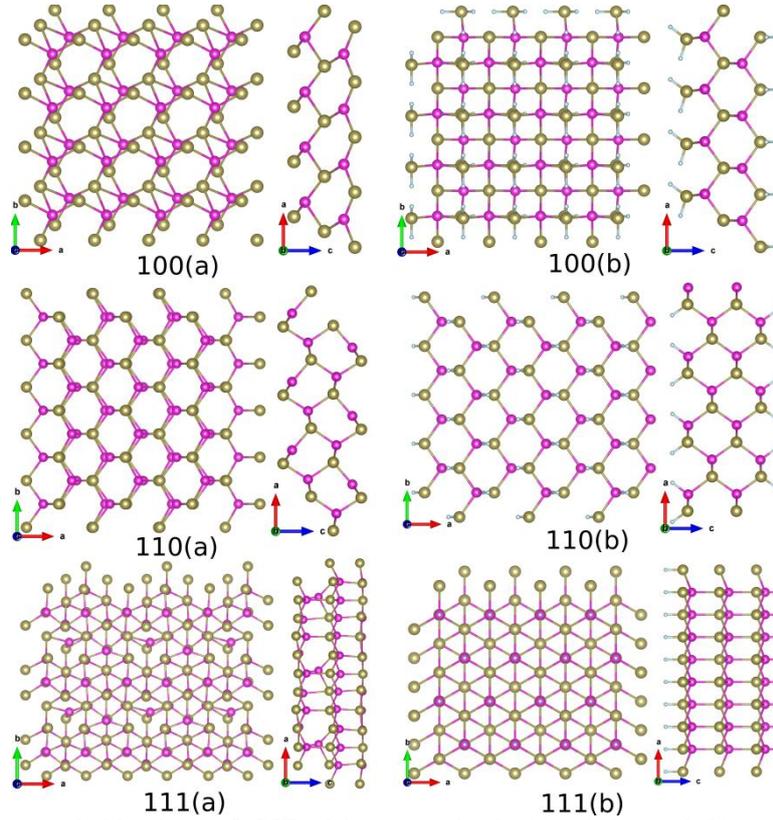

Figure 1. The top and side views of *CdTe* slab geometries in (a) pristine and (b) passivated form for different orientations.

The electronic band structure and density of states (DOS) for slabs along with their phonon dispersions, are shown in figure 2. The electronic band structures show that the structures are semiconducting with direct energy gap; except the pristine <100> slab which has indirect energy gap. The energy gap for pristine cases of <100> and <111> is 0.8 eV and 0.6 eV, respectively, which doubles on passivation and in <110> case, it widens from 1.3 eV to 1.4 eV.

The phonon band structure for *CdTe* slabs shows that all the modes at the zone center, i.e., the Γ point of Brillouin zone (BZ) have real frequencies, thus, these slab structures are dynamically stable. However, for the pristine <100> and <111> cases, an acoustic mode has the imaginary frequency in the Γ-X direction. The phonon PDOS has equal contributions due to *Cd* and *Te* till 50 cm$^{-1}$ except for the passivated

<100> and pristine <110> cases. In passivated <100> case, the *Te* and in pristine <110> case, the *Cd* have higher contribution to PDOS till 50 cm$^{-1}$ for optical modes. In pristine <100> case, the contribution to phonon DOS till 50 cm$^{-1}$ is equally due to *Cd* and *Te*, whereas the latter dominates till 130 cm$^{-1}$ and in the passivated case in close vicinity of 100 cm$^{-1}$. The PDOS, beyond these frequencies, has equal contribution due to *Cd* and *Te* and the states above 150 cm$^{-1}$ are due to FHPAs.

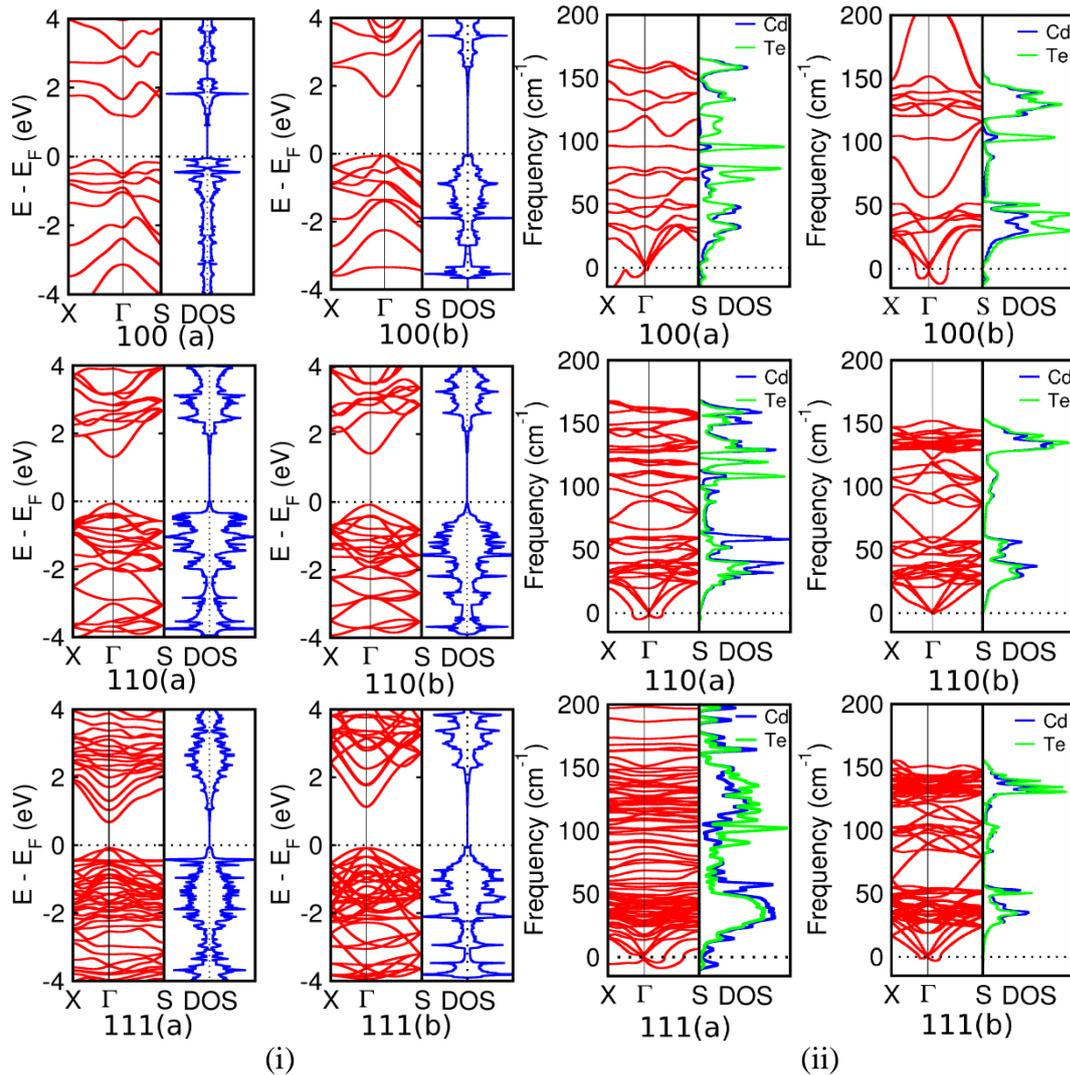

Figure 2. (i) The electronic band structure and DOS, (ii) phonon dispersions with PDOS for (a) pristine and (b) passivated form of *CdTe* slab in different orientations.

In the pristine <110> case, the phonon DOS has *Cd* contribution dominating till 60 cm$^{-1}$. Whereas the *Te* contribution is more between 100 cm$^{-1}$ to 130 cm$^{-1}$ and for the rest of the frequencies, the *Cd* and *Te* have equal contributions. In the passivated case, contribution to phonon DOS due to *Cd* and *Te* is equal. The states due to FHPAs lie at higher frequencies. In pristine <111> case, the phonon PDOS has slightly higher contribution due to *Cd* till the 70 cm$^{-1}$ and above that the *Te* dominates till 150 cm$^{-1}$. In the passivated case, the contribution to phonon modes is equal due to *Cd* and *Te* till 120 cm$^{-1}$; further, till 150 cm$^{-1}$ the contribution due to *Te* is more. The initial optical modes of vibration in *CdTe* slabs are due to the shear-like in-plane motions of the atoms.

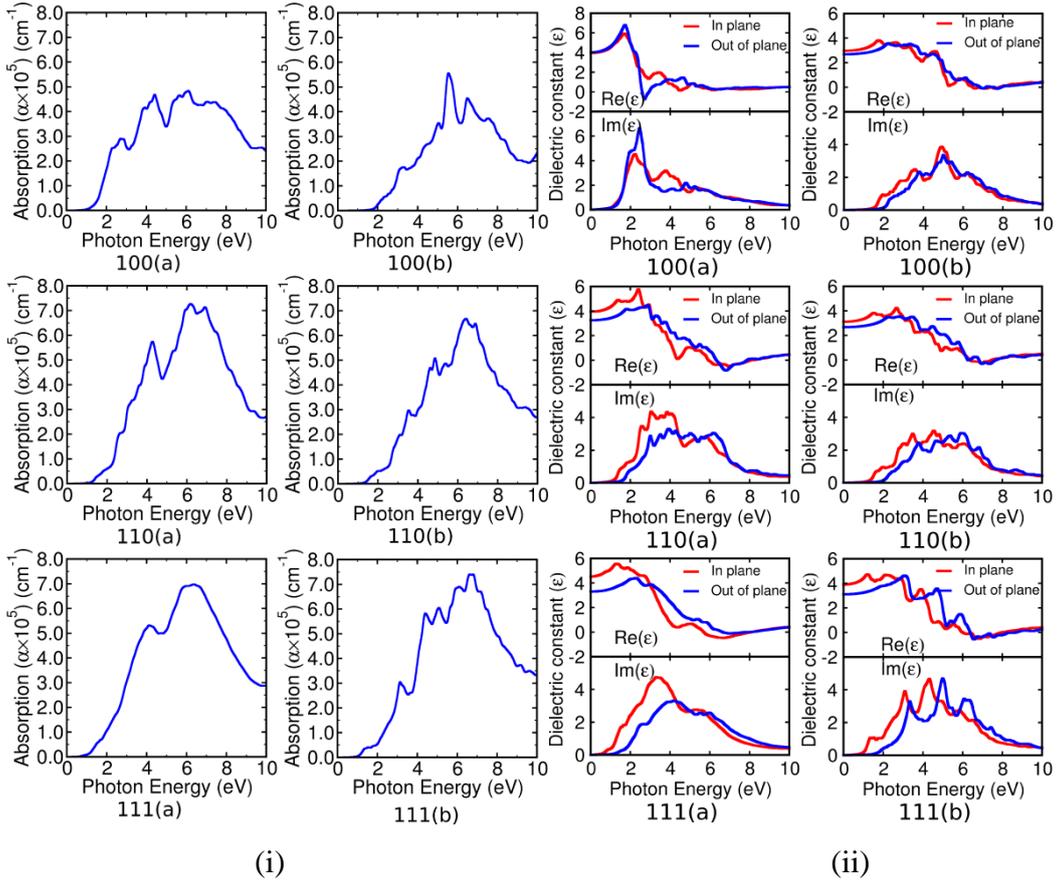

Figure 3. (i) The absorption coefficient and (ii) real and imaginary part of dielectric constant of *CdTe* slab in (a) pristine and (b) passivated form for different orientations. The absorption coefficient is plotted for in-plane direction.

The (i) absorption coefficient and the (ii) dielectric constants for *CdTe* slabs are shown in figure 3. The dielectric constants are plotted for the in-plane and out-of-plane directions, whereas the absorption coefficients are plotted for the in-plane direction. The dielectric constants in different directions are consistent qualitatively. The in-plane dielectric constant's real part dominates till 3 eV and imaginary part dominates till 4 eV over that of out-of-plane case. Beyond that, the dielectric constant for out-of-plane dominates till the 7 eV and for higher energies, dielectric constants overlap for in-plane and out-of-plane directions. However, in the pristine <100> case, the in-plane and out-of-plane dielectric constants are almost equal for 0 eV to 2 eV and above 5 eV. In the incident photon energy range of 2 eV to 5 eV both dielectric constant curves in two directions cross each other.

The frequency dependent optical spectra like absorption coefficient, reflectivity, refractive index, extinction coefficient, and energy loss functions can be calculated from the real and imaginary part of dielectric functions [39]. The absorption coefficients of slabs are near to $7\times10^5$ cm$^{-1}$ and peaks near the incident photon energy 6 eV. Although the absorption coefficient peak is in UV region, the edge of absorption spectrum lies in the IR, consistent with the energy gap of structures. Therefore, the thin slab of *CdTe* can absorb radiations in IR, visible and UV region of light spectrum.

**3.2 MONOLAYERS**

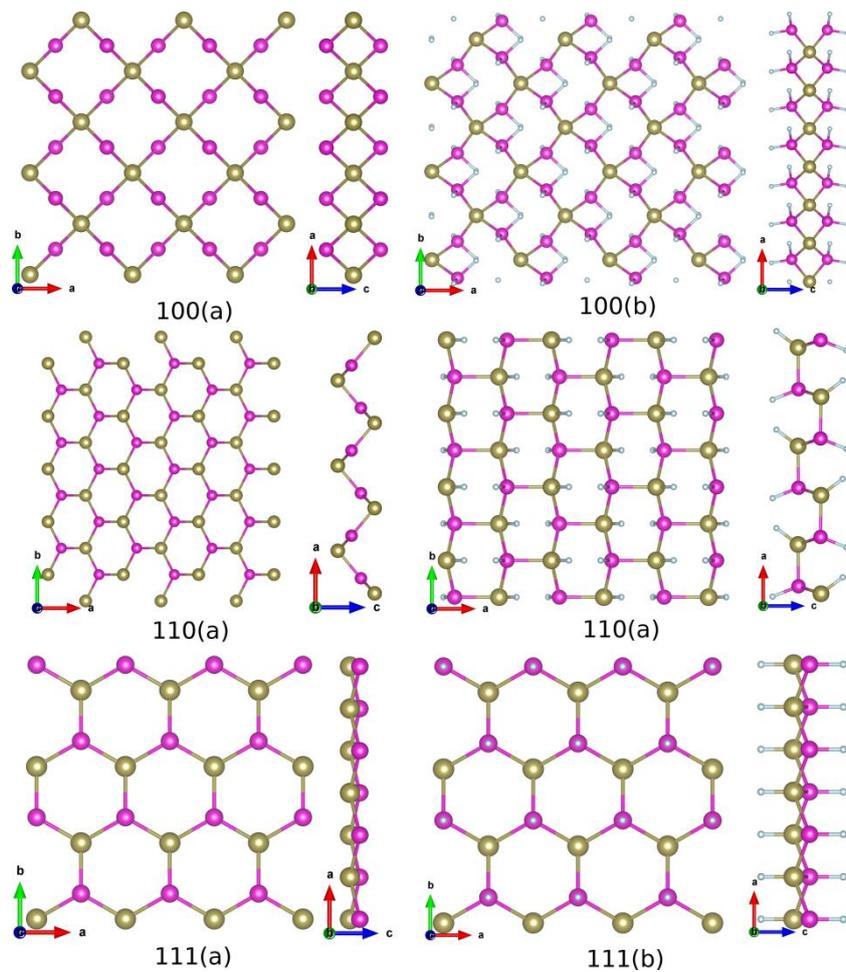

Figure 4. The top and side views of *CdTe* monolayer geometries in (a) pristine and (b) passivated form for different orientations.

Further to increase the confinement in *CdTe* 2D layers, the monolayers of *CdTe* in <100>, <110>, and <111> orientations have been studied and their geometries are shown in figure 4. The <100> monolayer has tetragonally bonded Tellurium layer with a square-like top view, <110> has the zig-zag hexagonal, and <111> has the planer hexagonal structure. On passivation, <100> retains its shape in the lateral view, but the top view is distorted as the *Te* atom in center is absent. The lateral view of <110> monolayer changes from zig-zag to hexagonal; however, in top view the hexagons appear as if they are elongated along the *b* direction; whereas, <111> monolayer remains planer with *Cd* and *Te* atoms moving outwards in the opposite direction. The thickness or width of these monolayers changes upon passivation. The thickness has reduced for the monolayer <100> (3.56 Å to 3.05 Å) and <110> (2.84 Å to 1.50 Å) and slightly increased for <111> (0.54 Å to 0.86 Å) as expected. The thickness of pristine <100> monolayer has increased from 3.49 Å to 3.56 Å after AIMD. The passivation changes *Cd-Te* bond lengths in monolayer, significantly. In <100> case the *Cd-Te* bond length lies in the range 2.73 Å - 2.92 Å which shortens to 2.80 Å - 2.84 Å and for <110>, it is from 2.78 Å - 2.82 Å to 2.81 Å - 2.83 Å. In the case of <111>, for pristine case *Cd-Te* bond length is 2.76 Å which increases on passivation to 2.84 Å. The *Cd-Te* bond lengths in monolayers are slightly smaller than that in the bulk.

The AIMD (figure S2) shows that the monolayers are stable at 300 K except for the pristine <100> monolayer where it transforms to another lower energy configuration at 300 K. Therefore, the final geometry of AIMD for pristine structure is quenched to 0 K and relaxed. This newly relaxed pristine <100> structure is used for further calculations. The change in the geometry of monolayer in <100> after AIMD is shown in figure S3.

The electronic band structure and DOS (figures 5(i)) shows that the monolayers are semiconducting with direct band gap. For the <100> case, the energy gap is 0.2 eV whereas, upon passivation, it widens to 3.0 eV. In this particular case, this energy gap does not occur at Γ point but at zone boundary. In the <110> and <111> the energy gap widens from 1.8 eV to 2.0 eV and 1.2 eV to 1.8 eV upon passivation, respectively. The energy gap of passivated cases and pristine <110> monolayer lies in visible spectrum.

The phonon dispersions and corresponding PDOS for the monolayers are shown in figures 5(ii). The monolayers have all the modes of vibrations with real frequency at Γ point except for the pristine <111> case, which has some optical modes with imaginary frequencies. The contribution to these imaginary states is predominantly due to the *Te* atoms. This is consistent with the earlier reports [7,18], however, the passivated hexagonal structure is dynamically stable. For the pristine <100> case, the couple of acoustic modes have an imaginary frequency as one goes away from the Γ point. Therefore except for pristine <111>, these structures are stable to lattice vibrations.

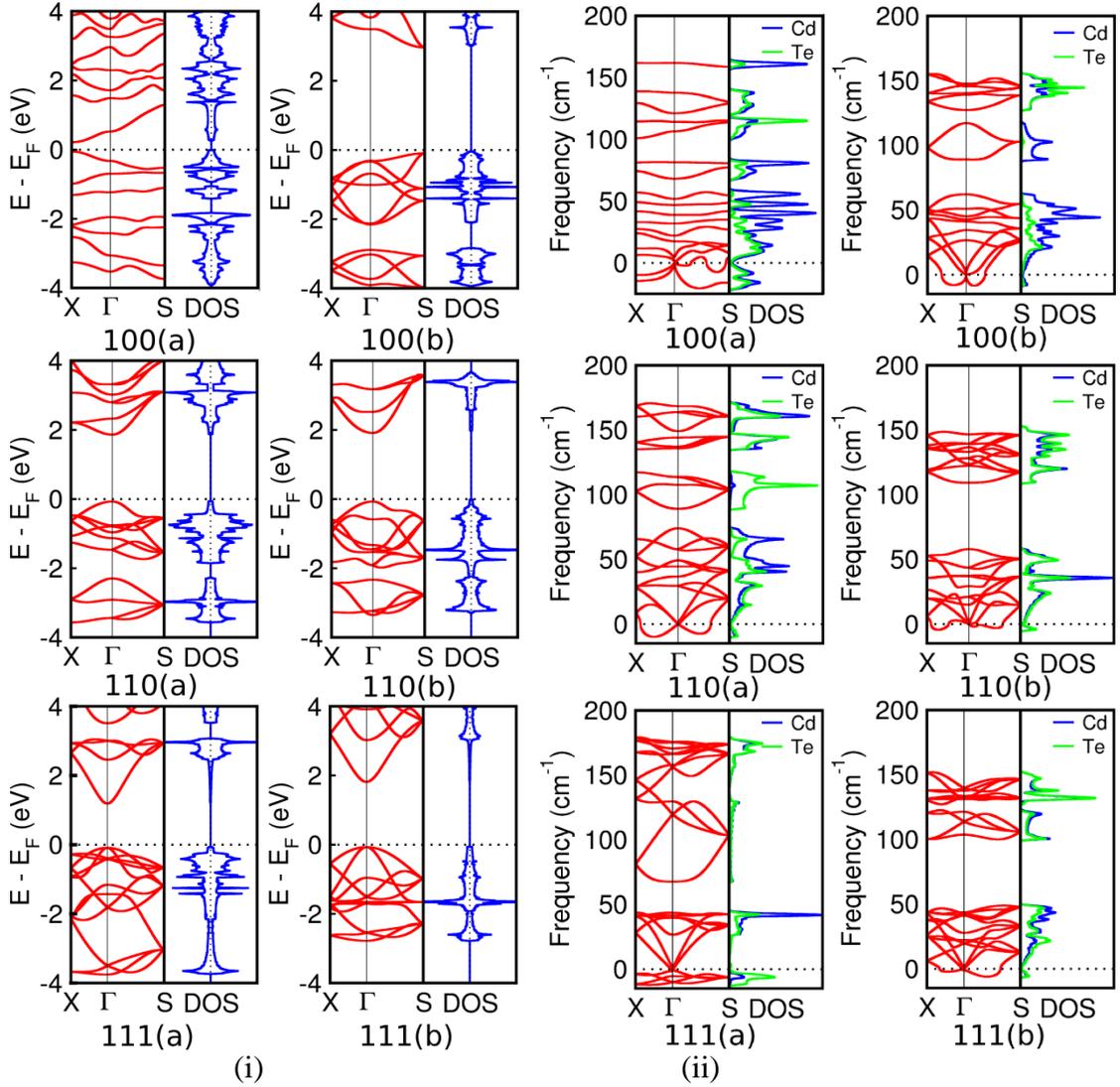

Figure 5. (i) The electronic band structure and DOS, (ii) phonon dispersions with PDOS of *CdTe* monolayer in (a) pristine and (b) passivated form for different orientations.

The phonon PDOS in pristine <100> and <110> have higher contribution due to *Cd* for the initial optical modes till nearly 80 cm$^{-1}$ and above 150 cm$^{-1}$ whereas for in between frequency, *Te* dominates. For passivated <100> case the *Cd* dominance is till

120 cm$^{-1}$ and beyond which *Te* contribution is higher. In pristine <110> case, the phonon DOS has a significant contribution due to *Cd* in the 40 cm$^{-1}$ to 80 cm$^{-1}$, while in the 80 cm$^{-1}$ to 120 cm$^{-1}$ range, it is solely due to *Te*. Beyond that, the states have almost equal contributions due to *Cd* and *Te*. In passivated structures, the phonon DOS has equal contribution due to *Cd* and *Te* except the <100> case where *Cd* dominates in lower frequencies and *Te* dominates in higher frequencies. The states due to FHPA lie at further higher frequencies.

The absorption coefficient and the dielectric constants of the monolayers are shown in figure 6. The dielectric constants are plotted for in-plane and out-of-plane directions, whereas the absorption coefficients are plotted for the in-plane direction of monolayers. The dielectric constants in the two directions differ significantly.

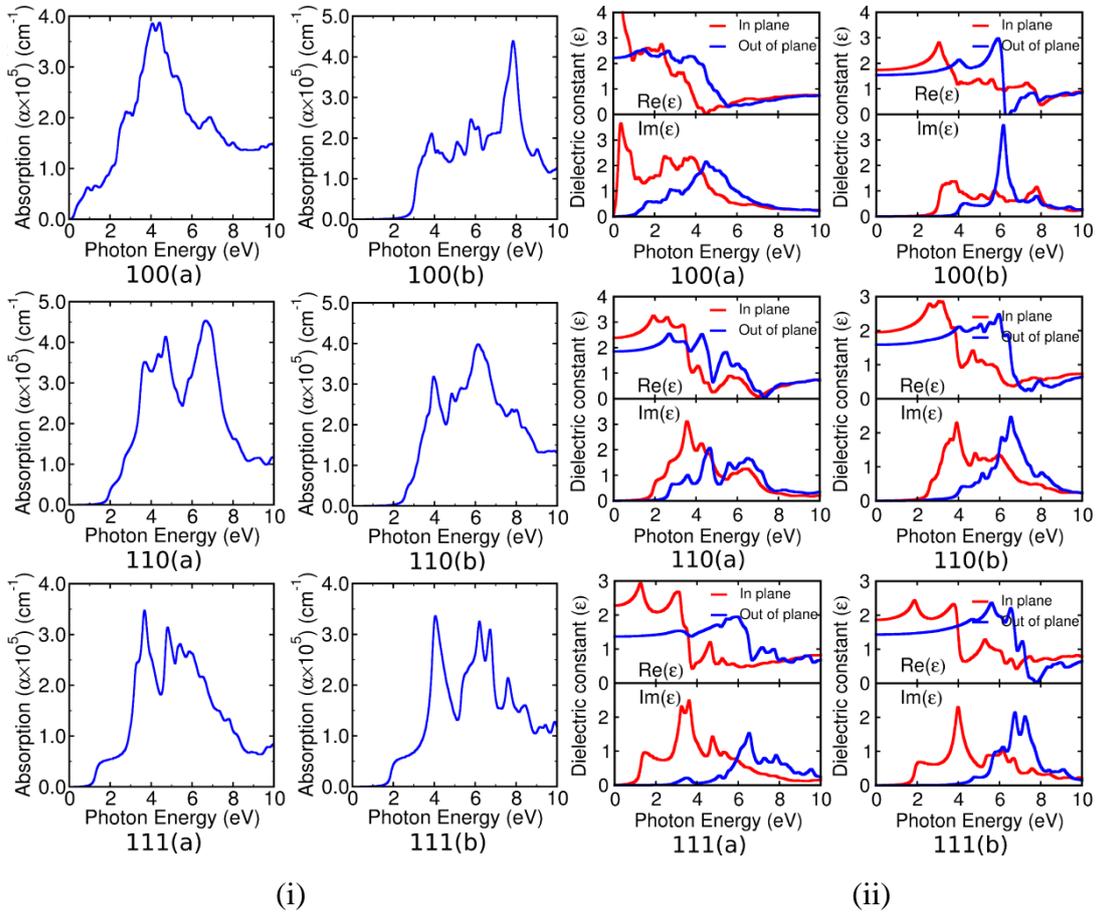

Figure 6. (i) The absorption coefficient and (ii) real and imaginary part of dielectric constant of *CdTe* monolayer in (a) pristine and (b) passivated form for different orientations. The absorption coefficient is plotted for in-plane direction.

The real part of the dielectric constant for the in-plane direction dominates roughly up to 4 eV of incident photon energy, beyond which the out-of-plane dielectric constant dominates. Beyond the 8 eV, the real part of the dielectric constant is independent of direction. The dielectric constant's imaginary part peaks at different incident photon energy energies. Upto specific incident photon energy (4 eV - 6 eV), the imaginary part for the in-plane direction dominates, and beyond that, the out-of-plane part dominates till 8 eV. In comparison with the slabs, the imaginary part of dielectric constant peaks at lower energies for the in plane direction whereas it peaks at higher energies for the out of plane direction. The absorption coefficient for the monolayers is almost equal to that of slabs and peaks around energy 4 eV and 6 eV. Similar to the slabs, the absorption edge of passivated monolayers and pristine monolayer in <110> orientation lies in the visible part of spectrum.

**3.3 RODS**

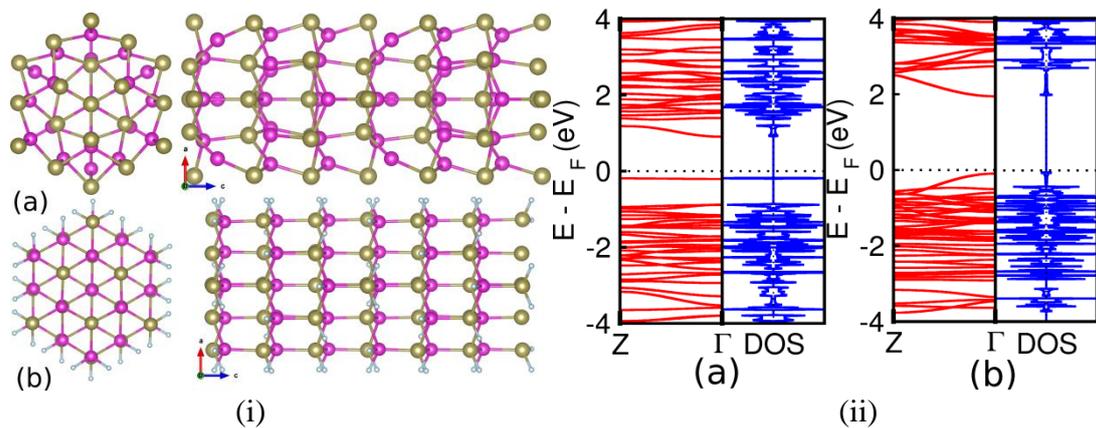

Figure 7. (i) The top and side views geometry and (ii) electronic band structure and DOS of (a) pristine and (b) passivated form of *CdTe* nanorod.

The *CdTe* nanorods are studied in <111> orientation. The geometry and electronic structures are shown in figure 7. The rod <111> has a round shape. In <111>, the nanorod has a round shape which becomes hexagonal on passivation. The <111> nanorod can be seen as the stacks of (111) monolayers. The nanorod in <111> orientation is stable at 300 K, as learned from the AIMD (Figure S4). As an effect of passivation, the diameter of the rod is observed to be reduced. In the <111>, the diameter reduces from 11.39 Å to 10.85 Å. The *CdTe* bond lengths in <111> case, change from 2.65 Å - 2.98 Å to 2.80 Å - 2.83 Å on passivation. The direct energy gap of 0.9 eV in pristine <111> widens up to 2.0 eV on passivation.

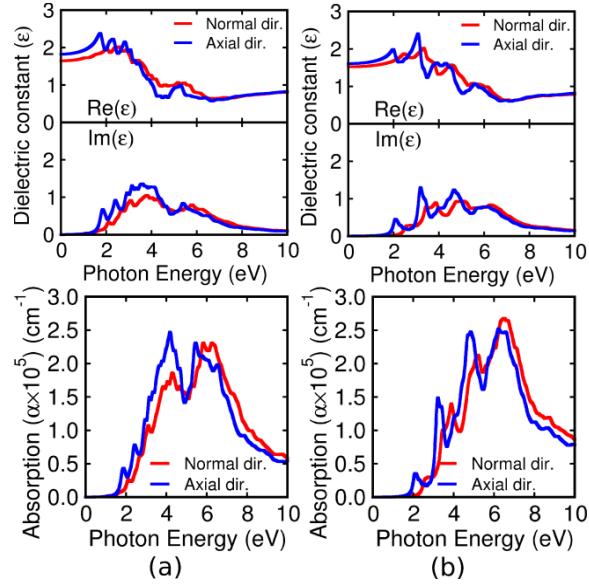

Figure 8. The real and imaginary part of dielectric constant (top) and the absorption coefficient (bottom) of *CdTe* nanorods in (a) pristine and (b) passivated form.

The phonon dispersion and corresponding partial DOS for rod in <111> orientations are shown in figure S5. For both, pristine as well as passivated nanorods are stable to lattice vibrations as all the modes have real frequencies at zone center. The pristine case has phonon PDOS with *Cd* dominance in 40 cm$^{-1}$ to 70 cm$^{-1}$ and *Te* dominance in 100 cm$^{-1}$ to 150 cm$^{-1}$ range. The contribution to phonon PDOS due to *Cd* and *Te* in passivated case is equal.

The absorption coefficient and dielectric constants for the axial direction and perpendicular direction to nanorod are shown in figure 8. The absorption coefficient and dielectric constants in nanorod's normal and axial directions do not differ much qualitatively or quantitatively. The peaks of absorption coefficient are at energy 4 eV and 6 eV however, the edge of passivated rod is around 2 eV which lies in the visible spectrum of light.

### 3.4 TUBES

The nanotubes of *CdTe* are derived from the various *CdTe* monolayers by rolling-up the rectangular monolayer. Out of those, three nanotubes are studied and their geometries and electronic structures are shown in figure 9. A tube is created from the <110> monolayer (Tube-I) and two nanotubes (Tube-II and Tube-III) are created from <111> monolayer by rolling it in two orthogonal directions. The top view of Tube-I and Tube-III is circular or round, whereas Tube-II has a hexagonal shape. The tube wall has a hexagonal structure, and *Te* atoms are on the nanotube surface.

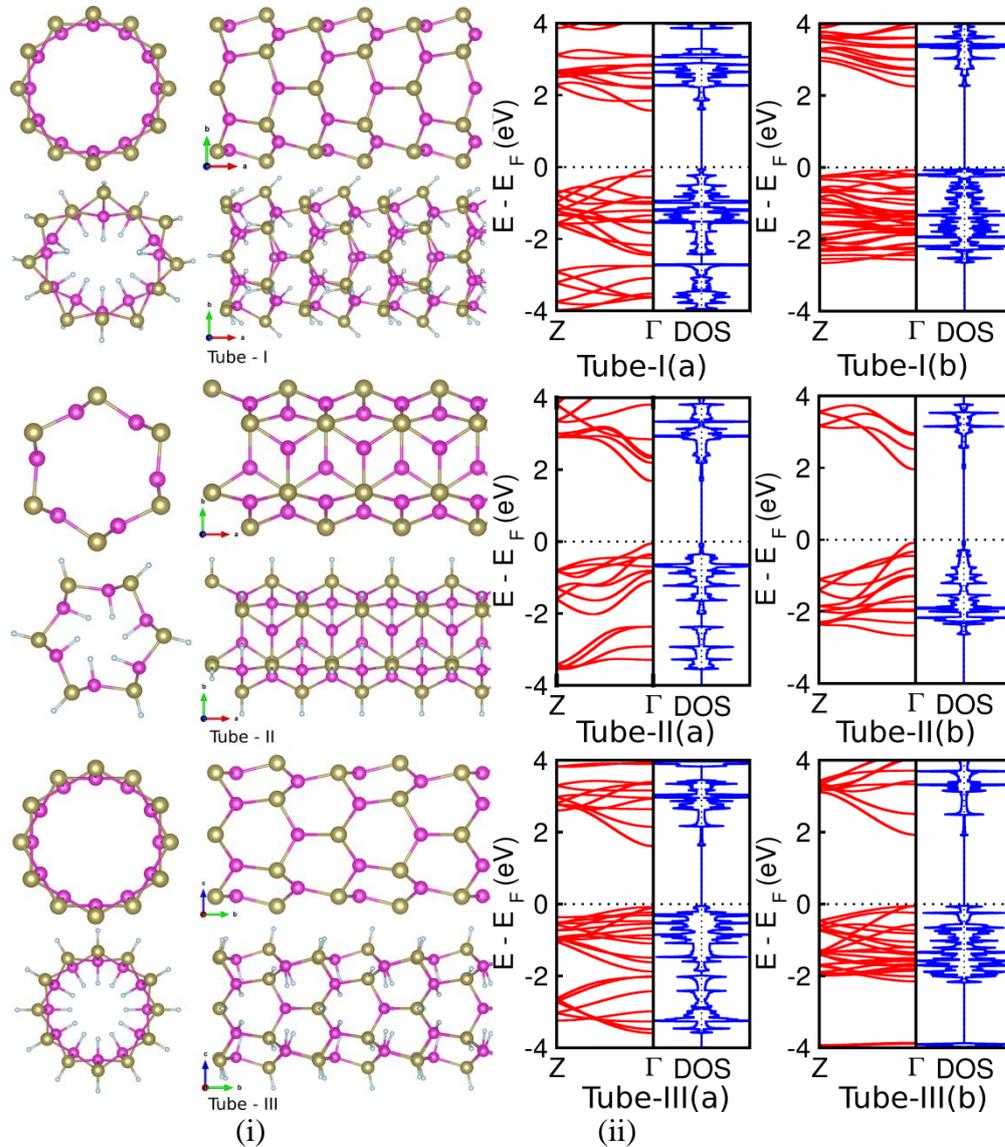

Figure 9. (i) The top and side views geometry and (ii) electronic band structure and DOS of (a) pristine and (b) passivated form of *CdTe* nanotubes.

When these nanotubes are characterized for their stability at 300 K, the pristine and passivated Tube-I loses its shape and form, whereas the pristine form of Tube-II and Tube-III gets distorted. The passivated nanotubes, Tube-II and Tube-III retain the shape and form at 300 K. The energy-time profiles at 300 K for the nanotubes are shown in figure S6.

The passivation has resulted in increase in the diameter of nanotubes and the changes are as follows, for Tube-I, it changes from 10.54 Å to 11.62 Å, for Tube-II it changes from 8.63 Å to 8.70 Å and for Tube-III, it changes from 9.71 Å to 10.03 Å. The *Cd-Te* bond length in Tube-I is 2.78 Å which changes upon passivation and lies between 2.79 Å - 2.98 Å. In Tube-II, it lies between 2.78 Å - 2.81 Å and changes to 2.82 Å - 2.85 Å; Tube-III, it has a range of 2.80 Å - 2.84 Å and becomes 2.86 Å - 2.90 Å on passivation. The passivation has resulted in to increase in *Cd-Te* bond length in nanotubes. There are relatively large variations in the bond lengths of passivated case in the Tube-I, while pristine has bond length of 2.78 Å for all *Cd-Te* bonds. All three tubes are semiconducting and their energy gap increases on passivation as expected. The energy gap for pristine Tube-I is 1.6 eV which increases to 2.2 eV on passivation; for Tube-II, it is from 1.7 eV to 2.0 eV; for Tube-III, it is from 1.6 eV to 2.0 eV. The energy gap values of pristine as well as the passivated nanotubes lie in the visible spectrum of light. It can be noticed that the diameter, energy gap has changed minimally for Tube-II among the nanotubes on passivation.

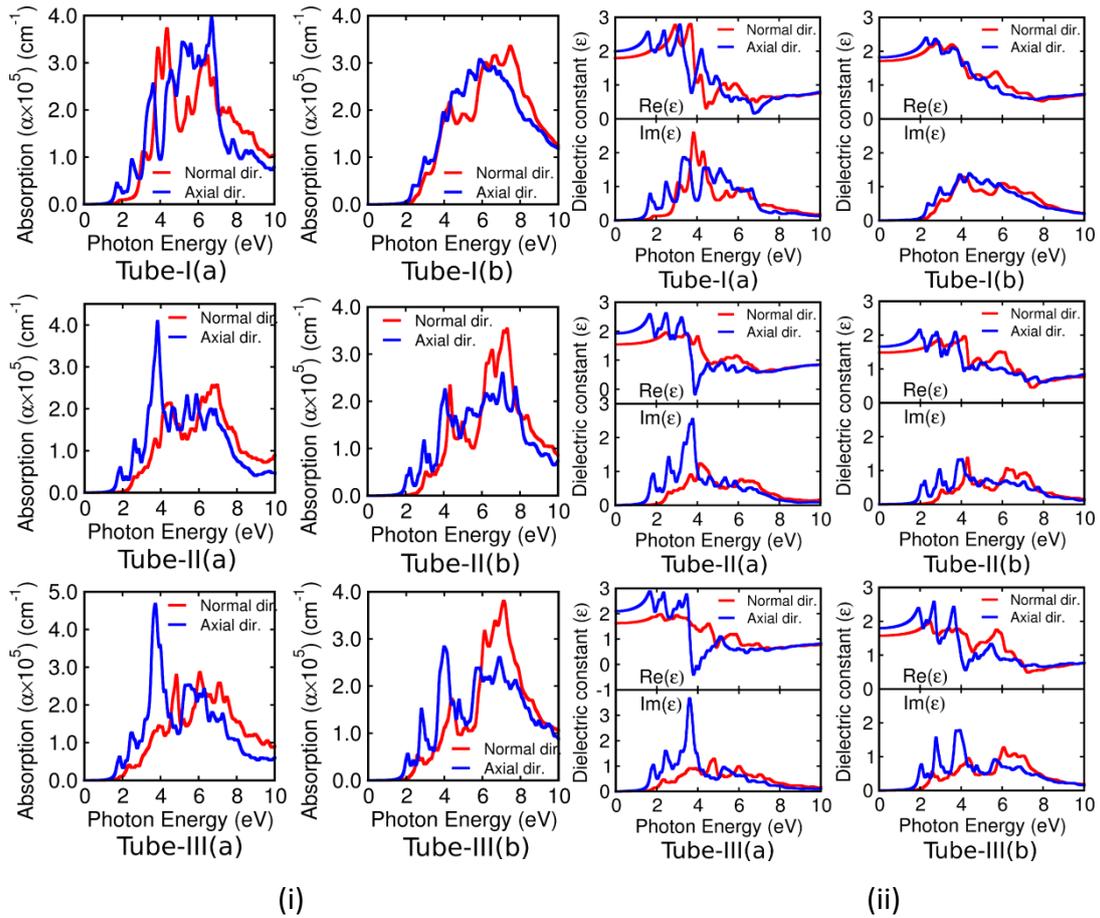

Figure 10. (i) The absorption coefficent and (ii) real and imaginary part of dielectric constant of *CdTe* nanotubes in (a) pristine and (b) passivated form.

The phonon dispersions and its PDOS for *CdTe* nanotubes are shown in figure S7. The pristine Tube-II, Tube-III and passivated Tube-I, Tube-II nanotubes have the phonon vibrational modes with real frequencies at the zone center and rest two cases have imaginary frequencies for certain modes at zone center. The contribution to phonon PDOS around the 50 cm$^{-1}$ is more due to the *Cd* whereas around 100 cm$^{-1}$ it is due to the *Te*. It can be seen from the phonon dispersions that the Tube-II in pristine as well as in passivated form is dynamically stable.

The absorption coefficient and dielectric constants along the axial and the normal directions of *CdTe* nanotubes are shown in figure 10. Unlike the nanorods, the dielectric constant differs along axial and normal to the tube direction. The dielectric constant in the axial direction is more significant than that in the normal direction. The absorption coefficient in axial direction peaks near 4 eV whereas the in normal/perpendicular direction it peaks near 7 eV. The absorption coefficient as well as the dielectric constant has oscillatory behavior along the axial direction.

| Structure | | | Thickness/ Diameter | Binding Energy (eV)/atom | Energy gap (eV) |
|---|---|---|---|---|---|
| SLAB | 100 | B | 5.14 | 1.8450 | 1.7 (0.8) |
| | | P | 6.55 | 1.9405 | 1.6 |
| | 110 | B | 7.49 | 1.9359 | 1.3 |
| | | P | 6.95 | 2.0364 | 1.4 |
| | 111 | B | 8.42 | 1.9036 | 0.6 |
| | | P | 8.51 | 2.0323 | 1.2 |
| MONO LAYER | 100 | B | 3.56 | 1.1702 | 0.2 |
| | | P | 3.06 | 2.1253 | 3.0 |
| | 110 | B | 2.84 | 1.8737 | 1.8 |
| | | P | 1.50 | 2.0049 | 2.0 |
| | 111 | B | 0.54 | 1.7839 | 1.2 |
| | | P | 0.86 | 2.0374 | 1.8 |
| ROD | 111 | B | 11.39 | 1.8189 | 0.9 |
| | | P | 10.85 | 2.0373 | 2.0 |
| TUBE | 110 | B | 10.55 | 1.8176 | 1.6 |
| | | P | 11.62 | 1.9720 | 2.2 |
| | 111-I | B | 8.63 | 1.8663 | 1.7 |
| | | P | 8.70 | 2.0011 | 2.0 |
| | 111-II | B | 9.71 | 1.8468 | 1.6 |
| | | P | 10.03 | 1.9948 | 2.0 |

Table 1. The thickness or diameter, binding energy per atom and direct energy gap of pristine (B) and passivated (P) *CdTe* nanostructures in different structural forms. The indirect energy gaps (if any) are mentioned in bracket.

The Coulomb repulsion among the surface *Te* atoms and the dangling bonds considerably determines the geometries of *CdTe* nanostructures. As a result, when the valance of a surface atom is completed, i.e., dangling bonds are satisfied, the nanostructure's geometry changes significantly concerning its pristine form. These changes in geometry are reflected in the thickness or diameter of the nanostructure and the *CdTe* bond lengths. The *CdTe* slab thickness has increased for the <100> and <111>, whereas it has decreased for <110> on passivation. In monolayers, the thickness reduces for <100> and <110> orientations, whereas it slightly increases for <111> case. In 1D systems, nanorods diameter decreases, whereas, for nanotubes, it increases on passivation. The *Cd-Te* bond lengths in passivated slabs and nanotubes are more or less uniform and are higher than the bond lengths in bulk *CdTe*; however, in monolayers and nanorods, the bond lengths are little less than that of bulk.

The energy band gap for slab along <100> and <111> doubles while for <110>, it merely widens by 0.1 eV on passivation. Consistent with the slab, the energy gap of monolayers <110> also shows a minimal change of 0.2 eV. However, in the <100> case, the energy gap drastically increases from 0.2 eV to 3.0 eV due to passivation, revealing the pristine <100> monolayer's instability. The solid 1D objects, *CdTe* pristine nanorod in <111> orientation, have an energy gap of 0.9 eV, increasing on passivation to 2.0 eV. In nanotubes, the energy gap in the pristine cases is around 1.6 eV, whereas, in the passivated form it is about 2.0 eV. Thus energy gaps for nanotubes increase by around 20 %-30 % upon passivation.

The AIMD at 300 K shows most of the structures, 2D and 1D, are stable at room temperature except the Tube-I, which is observed to be lost its form. The 2D systems like slab and monolayers in <110> orientation and 1D system Tube-III are the most structurally stable configurations at 300 K in pristine and passivated forms. Various optical properties can be derived from the dielectric constants. The nature of dielectric constants' real and imaginary parts gives an idea about the optical properties. The real part of the dielectric constant of slabs has a peak of around 2 eV, and for the imaginary part, it is around 4 eV. The absorption coefficient is relatively high (~$10^5$ cm$^{-1}$) and peaks around 6 eV. The energy gap of the slab increases on the passivation and lies in the near IR region; therefore, these structures can absorb light in a broad spectrum. The peak positions of dielectric constants in monolayers differ for in-plane and out-of-plane directions. For the in-plane direction, the dielectric constant peaks at lower energy than in the out-of-plane direction. The absorption coefficient in the monolayer is relatively less than slab, and its peak is around 4 eV and 6 eV. The energy gap of passivated monolayers is direct and around 2 eV; therefore, it can be blind to the IR and absorb the visible and UV light. In the *CdTe* nanorod, the dielectric properties are independent of directions, either along the axis or perpendicular. The peaks in the real and imaginary parts of its dielectric constants are observed to shift by 1 eV to higher energies on passivation. The absorption coefficient is of the order of slab or monolayer but lesser in magnitude. The absorption coefficient peak at 4 eV and 6 eV, and the energy gap of the pristine nanorod corresponds to the IR spectrum, whereas the energy gap for passivated nanorod corresponds to the visible spectrum and is thus blind to IR. Similar to other *CdTe* nanostructures, the peaks in the real and imaginary part of dielectric constants shift to higher energy on passivation in nanotubes. The dielectric constants in the axial direction are dominant for Tube-II and Tube-III. The *CdTe* nanotubes have an absorption edge in the visible light spectrum and therefore are blind to IR.

The thin slab, monolayers, and nanorods of *CdTe* are dynamically and structurally (at 300 K) stable, especially on passivation. As a result of stronger confinement in the monolayer than in the slabs, its energy gaps are more prominent and lie in the visible spectrum of light. The *CdTe* nanorod is stable, and its energy gap corresponds to the far IR region, which shifts to the visible region on passivation. The energy gap values of *CdTe* single-walled nanotube lie in the visible light spectrum and its independent of the passivation. Thus the single-walled *CdTe* nanotube's absorption edge lies in the visible spectrum; therefore, they are blind to IR. The minimal change in various properties of <110> slab and monolayer on passivation suggests that the structure has significantly self-healed. Although the energy gap changes almost equally in nanotubes on passivation, the stability at 300 K is better in Tube-III but not dynamically, in contrast to this, Tube-II is dynamically stable. Overall, for the stability of nanostructures, it needs to be passivated except for the <110> directed 2D system or nanotube Tube-III. The 2D system like <110> and nanotubes can be helpful for device applications due to their remarkable self-healing ability. The optical properties suggest that these nanostructures can have applications in optoelectronics, specifically solar cells.

## 4. Conclusions

The *CdTe* nanostructures such as thin slab, monolayer, nanorod, and nanotubes are investigated for electronic structure, room temperature stability, lattice vibrations, optical properties, and the effect on these properties due to passivation by FHPA. All the *CdTe* nanostructures studied are semiconducting with a direct energy gap which increases on passivation. The increase in the energy gap is minimal for the thin layer and monolayer of *CdTe* in <110> orientation and nanotubes, revealing the self-healing took place in their pristine structures. Such structures are essential for device applications. Most of the nanostructures (except Tube-I) are structurally stable at 300 K and their stability further increases on passivation. The optical properties show that the thin slab, nanorod, and pristine monolayers of *CdTe* can absorb the light in a broad spectrum of light (IR, Visible, and UV), whereas the monolayer on passivation becomes IR blind. In the nanotubes, the energy gap belongs to the visible region, even in pristine structures, and it increases on passivation by nearly 0.5 eV. As suggested by the thermal and dynamical stability study, such monolayers or single-walled nanotubes of *CdTe* can be stable, and passivation can enhance their stability further. Due to their energy gap and optical properties, these can be potentials candidate for optoelectronics or as photocatalysts for water splitting applications. The specific nanostructures show significant self-healing, which can be used for making the device and for spintronics applications.

## Acknowledgment

KTC acknowledges DAE/IGCAR for the research fellowship.

# Supporting Information

# Self-healing in unpassivated and passivated CdTe nanostructures: structural stability and optical properties

Kashinath T Chavan, Sharat Chandra

*Materials Science Group, Indira Gandhi Centre for Atomic Research, Kalpakkam 603102, Tamil Nadu, India*
*Homi Bhabha National Institute, Training School Complex, Anushakti Nagar, Mumbai, 400094, India*
email: ktchavan99@gmail.com, sharat@igcar.gov.in

| System | | Kpoint Grid |
|---|---|---|
| **Slab** | **Pristine** | 12×12×1 |
| | **Passivated** | 11×11×1 |
| **Mono Layer** | **Pristine** | 15×15×1 |
| | **Passivated** | 9×9×1 |
| **Rod** | **Pristine** | 1×1×9 |
| | **Passivated** | 1×1×5 |
| **Tube** | **Pristine** | 1×1×15 |
| | **Passivated** | 1×1×12 |

Table ST1: The energy cutoff for plane waves, and the K mesh used for electronic structure calculations of different *CdTe* nanostructures in the different structural forms.

**SLABS**

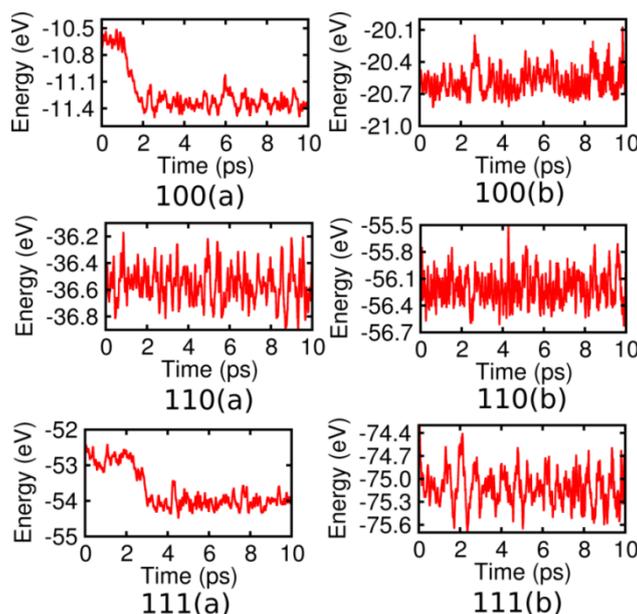

Figure S1. The total energy vs. time of AIMD simulation at 300 K of *CdTe* slab in (a) pristine and (b) passivated form for different orientations.

## MONOLAYERS

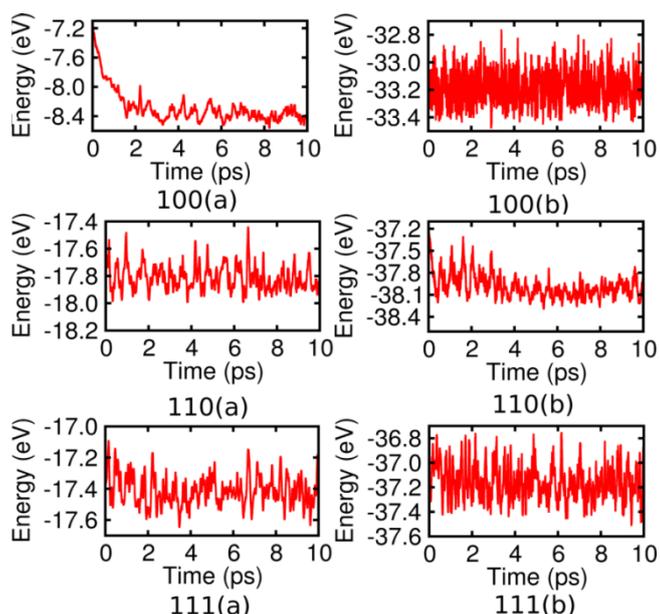

Figure S2. The total energy vs. time of AIMD simulation at 300 K of *CdTe* monolayer in (a) pristine and (b) passivated form for different orientations.

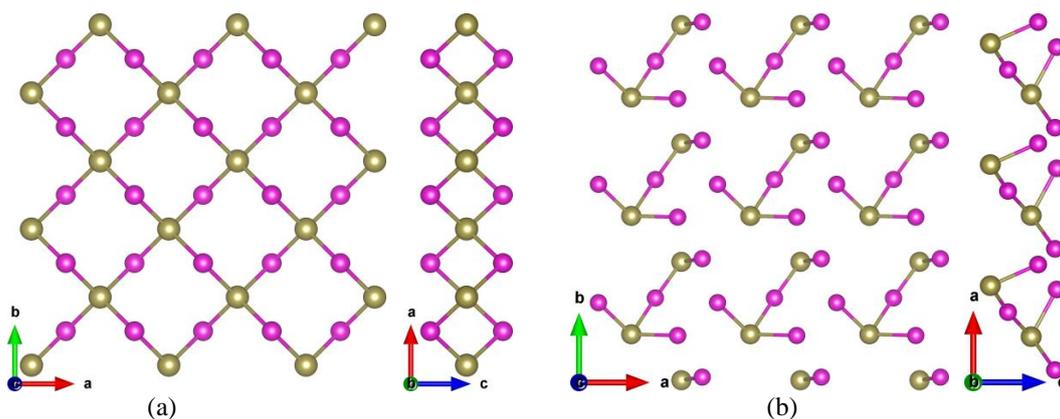

(a)          (b)

Figure S3. The top and side view of the initial geometry of monolayer in <100> direction and the one obtained from the thermalized monolayer at 300 after relaxation.

## RODS

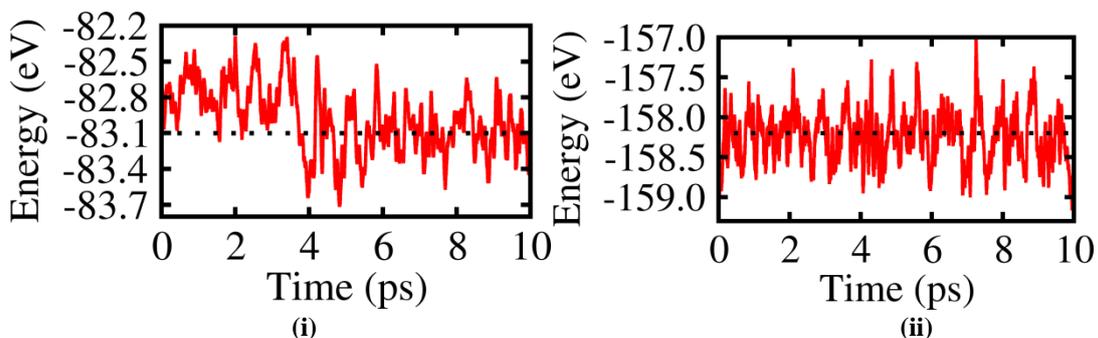

(i)          (ii)

Figure S4. The total energy vs. time of AIMD simulation at 300 K of CdTe nanorod in (i) pristine and (ii) passivated form.

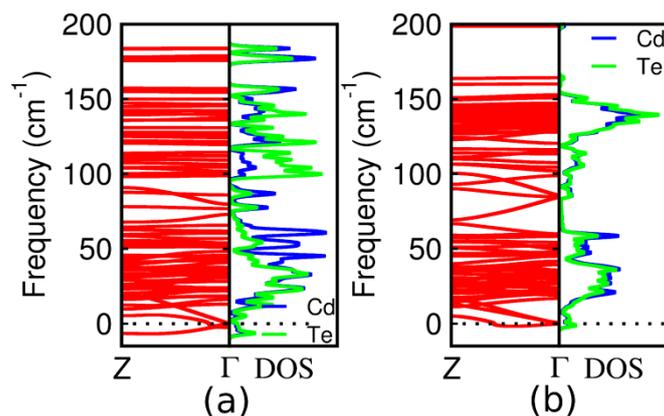

Figure S5. The phonon dispersions and corresponding partial DOS of CdTe nanorod in (a) pristine and (b) passivated form.

**TUBES**

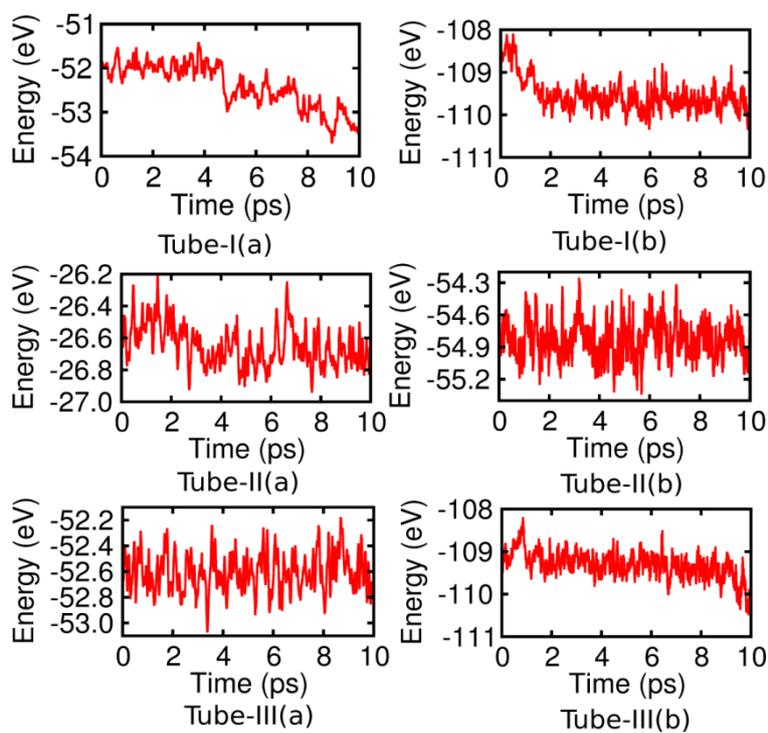

Figure S6. The total energy vs. time of AIMD simulation at 300 K of CdTe nanotubes in (a) pristine and (b) passivated form.

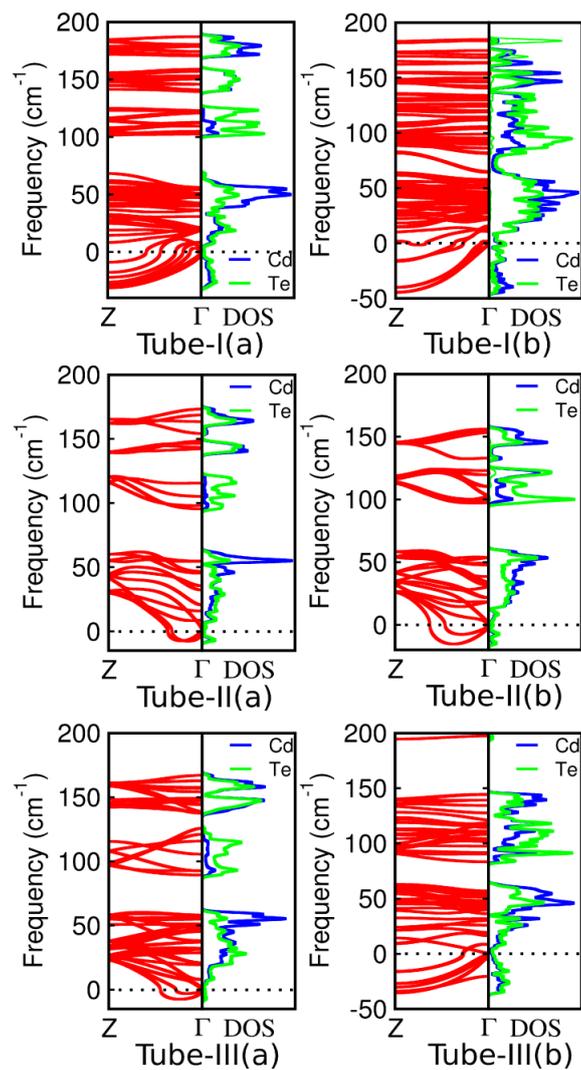

Figure S7. The phonon dispersions with PDOS of CdTe nanotube in (a) pristine and (b) passivated form for different orientations.